\documentclass[fleqn,10pt]{wlscirep}

\usepackage{times}
\usepackage{graphicx}
\usepackage{amsfonts}
\usepackage{amsmath, amsthm, amssymb}
\usepackage{dsfont}
\usepackage{mathptm}
\usepackage{color}

\newcommand{\eg}{\textit{e.g. }}
\newcommand{\etal}{\emph{et al.}}
\def\i{\mathrm{i}}

\title{Strong odd-frequency correlations in fully gapped Zeeman-split superconductors}

\author[1,*]{Jacob Linder}
\author[2]{Jason W. A. Robinson}
\affil[1]{Department of Physics, Norwegian University of
Science and Technology, N-7491 Trondheim, Norway}
\affil[*]{jacob.linder@ntnu.no}
\affil[2]{Department of Materials Science and Metallurgy, University of Cambridge, 27 Charles Babbage Road, Cambridge CB3 0FS, UK}

\begin{abstract}
It is now well established that at a superconductor/ferromagnet (S/F) interface an unconventional superconducting state arises in which the pairing state is odd-frequency. The hallmark signature of this superconducting state is generally understood to be an enhancement of the electronic density of states (DoS) at subgap energies close to the S/F interface. However, here we show that an odd frequency state can be present even if the DoS is fully gapped. As an example, we show that this is the case in the pioneering S/FI (where FI is a insulating ferromagnet) tunneling experiments of Meservey and Tedrow, and we derive a generalized analytical criterium to describe the effect of odd-frequency pairing on the DoS. Finally, we propose a simple experiment in which odd-frequency pairing in a Zeeman-split superconductor can be unambiguously detected via the application of an external magnetic field.
\end{abstract}
\begin{document}

\flushbottom
\maketitle

\thispagestyle{empty}

\section*{Introduction}

The search for unconventional superconductors which possess properties such as spin-polarization and topologically robust surface states, is currently of significant interest \cite{linder_nphys_15, qi_rmp_11}. Examples of such superconductors include uranium-based heavy fermions \cite{saxena, aoki} where ferromagnetism and superconductivity coexist; noncentrosymmetric superconductors \cite{bauer} with mixed singlet and triplet pair correlations \cite{gorkov}; and topological superconductors \cite{kane_review, qi_rmp_11} such as Bi$_2$Se$_3$ or hybrid topological insulator / superconductor structures, where Majorana zero-modes featuring non-Abelian exchange statistics are predicted \cite{fu, sau, alicea}. Although these are examples of very different materials systems, they share the common feature that the superconducting state cannot be fully described by the Bardeen-Cooper-Schrieffer theory of superconductivity \cite{bcs}. In fact, as will be explained below they are all closely related to the appearance of a so-called odd-frequency superconducting state.

An unconventional superconducting order in which the pairing state is odd-frequency was first speculated \cite{berizinskii} to arise in liquid $^3$He and was later predicted at $s$-wave superconductor/ferromagnet interfaces (S/F) by Bergeret \etal \cite{bergeret}. Subsequent studies demonstrated that odd-frequency pairing in fact appears in a wide variety of physical systems as a result of symmetry breaking \cite{eschrig_jlowtemp, tanaka} and that both Andreev-states bound to surfaces due to interface scattering and even Majorana fermions are accompanied by the emergence of odd-frequency pairing \cite{asano_prb_13}. These phenomena are central components in the above mentioned systems, which means that odd-frequency pairing is highly relevant in a wide range of superconducting systems ranging from heavy-fermion compounds to topological superconductors.  Moreover, odd-frequency superconductivity is very attractive for practical applications since it is robust against impurity scattering and the paramagnetic limitation. The hallmark signature of odd frequency pairing is commonly understood to be an enhancement of the density of states (DoS) \cite{braude} at low energies at S/F interfaces and although this property has been studied in a number of theoretical works \cite{theory_oddfreq}, experimental results are inconclusive \cite{kontos_prl_01, experiment_oddfreq}. Other properties of odd-frequency pairing, such as its effect on spin-susceptibility \cite{mizushima_prb_14}, its Meissner response \cite{meissner}, and even its appearance in Bose-Einstein condensates \cite{balatsky_bec}, topological insulators and graphene \cite{linderblack_prb_10, black_prb_12}, and multiband systems \cite{black_prb_13}, have recently garnered attention.

In this article, we show that an odd-frequency state can persist even when the DoS is fully gapped. This surprising finding has important consequences, most especially at a fundamental level where the identification of odd-frequency pairing via DoS experiments is an area of particular focus. Currently the best proof of triplet pairing is indirect and inferred from supercurrent or critical temperature measurements of S/F structures. Here we report three main results: first, we show that Zeeman-split superconductors \cite{meserveytedrow} are an example of a system in which odd-frequency pairing is present despite the fully gapped DoS; secondly, we derive a general analytical criterium for when odd-frequency pairing is present without causing an enhancement of the low-energy DoS; thirdly, we propose a straightforward experiment in which the existence of odd-frequency pairing in a Zeeman-split superconductor can be unambiguously demonstrated and in which spin-polarized triplet Cooper pairs can be generated via the application of a magnetic field. Since there is currently great interest in various forms of unconventional superconductivity as manifested in the systems mentioned above, where odd-frequency pairing will be present, our findings of how the odd-frequency superconducting state is experimentally manifested therefore bear upon a range of different physical systems. Specifically, our results are of importance with regard to the interpretation of the presence or absence of a zero-bias conductance peak, which is usually taken to be one of the hallmark signatures for unconventional superconducting pairing. 

\section*{Results}

\begin{figure}[t!]
\centering
\resizebox{0.45\textwidth}{!}{
\includegraphics{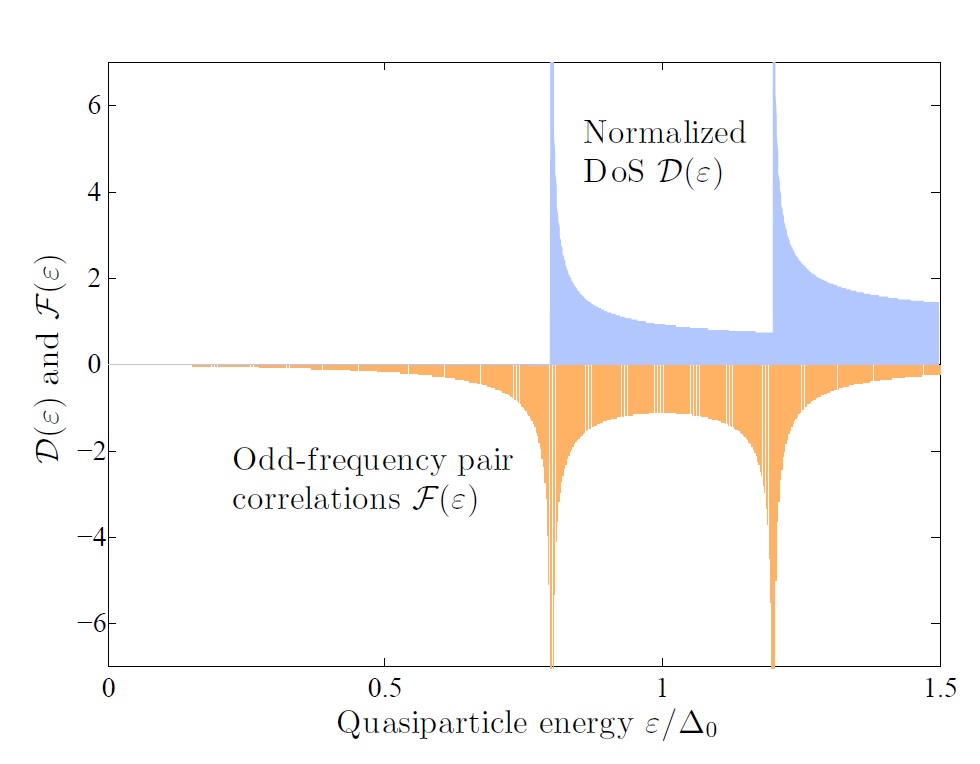}}
\caption{(Color online) Plot of the normalized density of states (DoS) $\mathcal{D}$ and the odd-frequency pairing amplitude $\mathcal{F} = |f_t|^2$ for a Zeeman-split superconductor with $h_S/\Delta_0=0.2$. The axis for $\mathcal{F}$ has been inverted in the plot to avoid overlap with the DoS. }
\label{fig:dos} 
\end{figure}

We demonstrate this effect in a simple model system: a Zeeman-split $s$-wave superconductor in the low-temperature limit where $T\ll T_c$.  This system may be realized in a thin-film superconductor with an in-plane applied magnetic field \cite{meserveytedrow}. We are thus first considering a single superconducting film in the presence of an external magnetic field, i.e. without any proximity effect to a ferromagnetic material. The induced exchange field splits the DoS for the spin-species which allows for the presence of a superconducting state below the Clogston-Chandrasekhar \cite{cc} limit $h/\Delta_0 < 1/\sqrt{2}$, where $\Delta_0$ is the superconducting gap at $T=0$. By considering the diffusive limit which is relevant experimentally, the Green's function $\hat{g}$ describing this setup may be computed using the quasiclassical theory \cite{quasiclassical}. The basic assumption underlying the quasiclassical approach is that the Fermi wavelength is by far the smallest length scale in the problem, or conversely that the Fermi energy is much larger than other energy scales such as the superconducting gap and the exchange field. It is known to give a very satisfactory description of the induced triplet pairing in superconducting systems \cite{buzdin_rmp_05}. To find the Green's function in the diffusive limit, we solve the Usadel \cite{usadel} equation:
\begin{align}\label{eq:usadel}
[\epsilon\hat{\rho}_3 + \hat{\Delta} + \hat{M}, \hat{g}]_- = 0.
\end{align}
Here, $[\ldots]_-$ denotes a commutator, $\varepsilon$ is the quasiparticle energy, $\Delta_0$ is the superconducting gap, $h_S$ is the induced exchange field, and the matrices inside the commutator are:
\begin{align}
\hat{\Delta} = \Delta_0\begin{pmatrix}
\underline{0} & \i\underline{\sigma}_y\\
\i\underline{\sigma_y} & \underline{0}\\
\end{pmatrix},\; 
\hat{M} =  h_S\begin{pmatrix}
\underline{\sigma}_z & \underline{0} \\
\underline{0} & \underline{\sigma}_z \\
\end{pmatrix},\; 
\hat{\rho}_3 = \begin{pmatrix}
\underline{1} & \underline{0} \\
\underline{0} & -\underline{1} \\
\end{pmatrix}
\end{align}
when using a real $U(1)$ gauge for the superconducting order parameter. The Green's function is required to satisfy normalization $\hat{g}^2 = \hat{1}$ and the solution of Eq. (\ref{eq:usadel}) is then obtained in the form:
\begin{align}
\hat{g} &= \begin{pmatrix}
c_\uparrow & 0 & 0 & s_\uparrow \\
0 & c_\downarrow & s_\downarrow & 0 \\
0 & -s_\downarrow & -c_\downarrow & 0 \\
-s_\uparrow & 0 & 0 & -c_\uparrow \\
\end{pmatrix}.
\end{align}
For easier analytical expressions, one may note that for positive energies $\varepsilon>0$ and exchange fields smaller than the gap, one has $c_\sigma = (\varepsilon + \sigma h_S)/\sqrt{(\varepsilon + \sigma h_S)^2 - \Delta_0^2}$ and $s_\sigma = \sigma\Delta_0/\sqrt{(\varepsilon +\sigma h_S)^2 - \Delta_0^2}.$ The expressions for $\{c_\sigma,s_\sigma\}$ valid for arbitrary energies and temperatures, in particular $T$ close to $T_c$ where $h_S$ may exceed $\Delta(T)$, are $c_\sigma=\text{cosh}(\theta_\sigma)$ and $s_\sigma = \text{sinh}(\theta_\sigma)$ with $\theta_\sigma = \text{atanh}(\sigma \Delta(T)/(E+\sigma h_S)$. In order to obtain information about the nature of the superconducting state, we make use of the definition of the singlet and triplet anomalous Green's functions which in our notation read $f_s = (s_\uparrow - s_\downarrow)/2$ and $f_t = (s_\uparrow + s_\downarrow)/2$. This allows us to write down the following explicit expression for the anomalous Green's functions, which in the limit $h_S\to 0$  reduces to the conventional BCS case with $f_t=0$:
\begin{align}\label{eq:anomalous}
f_{s(t)} = \frac{\Delta_0}{2}\{ [(\varepsilon+h_S)^2 - \Delta_0^2]^{-1/2} \pm [(\varepsilon-h_S)^2 - \Delta_0^2]^{-1/2} \}.
\end{align}
We emphasize that the triplet component is odd in frequency whereas the singlet component has a conventional BCS pairing correlation. In general, however, $f_t$ remains finite. Although this may seem paradoxical, STM measurements on Zeeman-split superconductors by Meservey and Tedrow \cite{meserveytedrow} demonstrated that the DoS remains fully gapped at low energies with a zero DoS whereas the BCS coherence peak is split in two. At the same time, the signature of odd-frequency pairing in superconductor/ferromagnet structures has previously been shown to be an enhancement of the low-energy DoS, both on the F and S side, resulting in some cases in large zero-bias conductance peaks \cite{braude, linder_prl_09}. From Eq. (\ref{eq:anomalous}) it is clear that $f_t$ is finite across the entire energy range $\varepsilon>0$ despite the fact that the normalized DoS $\mathcal{D}$ is equal to zero in the low-energy regime. In fact, $f_t$ grows substantially in magnitude as one approaches the excitation energies corresponding to the first spin-split peak $\varepsilon_h \equiv \Delta - h$, but the DoS remains zero. We illustrate this graphically in Fig. \ref{fig:dos} where the odd-frequency pairing amplitude is $\mathcal{F} = |f_t|^2$. This means that odd-frequency pairing does not necessarily result in an enhancement of the DoS as generally believed. 

\subsection*{Analytical criterium for enhanced subgap density of states due to odd-frequency pairing}

In order, therefore, to distinguish between the odd and even frequency components in the DoS of a Zeeman split superconductor, it is neccesary to enhance the spectral weight of the odd frequency component. We now proceed to show that the physics controlling this behavior lies in whether the singlet and triplet components are in-phase or out-of-phase. The DoS $\mathcal{D}$ may be computed in the standard way by taking the real part of the normal part of the Green's function, which in our notation gives $\mathcal{D}(\varepsilon) = \frac{1}{2} \sum_\sigma \text{Re}\{ c_\sigma\}.$ In order to establish a direct relation between the DoS and the anomalous Green's function, and thus the pairing symmetry realized in the superconductor, we use that $(c_\uparrow+c_\downarrow)( c_\uparrow-c_\downarrow) = c_\uparrow^2-c_\downarrow^2 = s_\uparrow^2-s_\downarrow^2$ due to the normalization condition for the components of the Green's function $\hat{g}$. As long as $c_\uparrow-c_\downarrow \neq 0$, meaning that the density of states is not spin-degenerate (which in particular is valid at all energies $\varepsilon$ for the system considered presently) we may then write:
\begin{align}
\sum_\sigma c_\sigma  &= \frac{s_\uparrow^2-s_\downarrow^2}{c_\uparrow-c_\downarrow} = \frac{(s_\uparrow+s_\downarrow)(s_\uparrow-s_\downarrow)}{c_\uparrow-c_\downarrow} = \frac{4f_t f_s}{(c_\uparrow-c_\downarrow)},
\end{align}
which produces the result $\mathcal{D}(\varepsilon) = 2\text{Re}\Big\{ \frac{f_t(\varepsilon)f_s(\varepsilon)}{c_\uparrow-c_\downarrow}\Big\}.$ From this equation, one may demonstrate how the presence of odd-frequency pairing $f_t$ in itself is \textit{not sufficient} in order to produce a finite DoS $\mathcal{D}(\varepsilon)$. Instead, a requirement is placed on the relative phase between the singlet and triplet pairing correlations in the system. To see this clearly, consider our specific system of a Zeeman-split superconductor. Using the previously derived expressions for $c_\sigma$, it is seen that $c_\sigma$ is purely imaginary below the first spin-split peak in Fig. \ref{fig:dos}, i.e. at $\varepsilon < \varepsilon_h$. From the fundamental definition of the DoS, this immediately implies that there should be no available quasiparticle states in this energy interval (as also seen in Fig. \ref{fig:dos}). Thus, the density of states is exactly zero in this energy regime. We are now in a position to show that this happens \textit{even in the presence of strong odd-frequency pairing}. Using that $c_\sigma$ has no real component (zero DoS), we obtain the final result for the density of states at energies $\varepsilon < \varepsilon_h$:
\begin{align}\label{eq:finalD}
\mathcal{D}(\varepsilon) = \frac{2\text{Im}\{f_t(\varepsilon)f_s(\varepsilon)\}}{\text{Im}\{c_\uparrow-c_\downarrow\}} = 0.
\end{align}
The above equation expresses a crucial piece of information:  when $f_t$ and $f_s$ are both real or both imaginary, hereafter referred to as in-phase, $\mathcal{D}(\varepsilon) = 0$ \textit{regardless of the magnitude of $f_t$}. Conversely, in order for the presence of odd-frequency pairing $f_t$ to yield an enhancement of the DoS, it needs to be out-of-phase with the singlet component $f_s$. We underline that Eq. (\ref{eq:finalD}) is valid when $\mathcal{D}(\varepsilon)=0$, demonstrating that the density of states vanishes when $f_t$ and $f_s$ are in-phase.

Our result is consistent with previous measurements on ferromagnet/superconductor (F/S) bilayer system \cite{kontos_prl_01}. In this case, odd-frequency correlations are also present and are known to be out-of-phase with the singlet component \cite{buzdin_rmp_05}. This leads to an alternating enhancement and suppression of $\mathcal{D}(\varepsilon)$ as one varies the junction width $L$, since the relative phase varies with $L$. The new insight gained from the present derivation is that odd-frequency pairing in itself is not enough to enhance $\mathcal{D}(\varepsilon)$ which has been believed to be a smoking gun signature for its existence:  the reason odd-frequency pairing provides such an enhancement in the F/S case is because it is out-of-phase with the singlet component. In fact, it is this scenario that has been investigated in previous works that have found that odd-frequency pairing enhances the DoS \cite{braude, theory_oddfreq} such as superconductor/ferromagnet bilayers, which is why the effect predicted in this paper has not been discovered previously. In contrast, we have shown that the singlet and triplet Cooper pair wavefunctions can be in-phase in which case the DoS remains fully gapped even if odd-frequency is finite and strong in magnitude.

It should be noted that Eq. (\ref{eq:finalD}) does not imply that singlet pairing $f_s$ must be present in general for odd-frequency pairing $f_t$ to influence the DoS. For instance, it has been shown in Ref. \cite{linder_prl_09} that it is possible to induce pure odd-frequency pairing, without any singlet pairing, at the Fermi level $\varepsilon=0$ in a normal metal/superconductor bilayer with a spin-active interface region. This may be realized by using for instance a ferromagnetic insulator as a barrier separating the two layers. In that case, the odd-frequency component gives rise to a large zero-bias conductance peak despite the absence of singlet pairing. Nevertheless, this is consistent with the present result as the $c_\uparrow-c_\downarrow = 0$  for that system, in which case Eq. (\ref{eq:finalD}) is not valid. This shows that odd-frequency pairing can in itself (without singlet pairing) enhance the low-energy DoS, but it does not necessarily do so. In particular, we have shown here that for another system with coexisting superconducting and magnetic order the DoS remains fully gapped even in the presence of strong odd-frequency correlations.

\subsection*{Self-consistent calculations and proposal for experiment}

\begin{figure}[t!]
\centering
\resizebox{0.45\textwidth}{!}{
\includegraphics{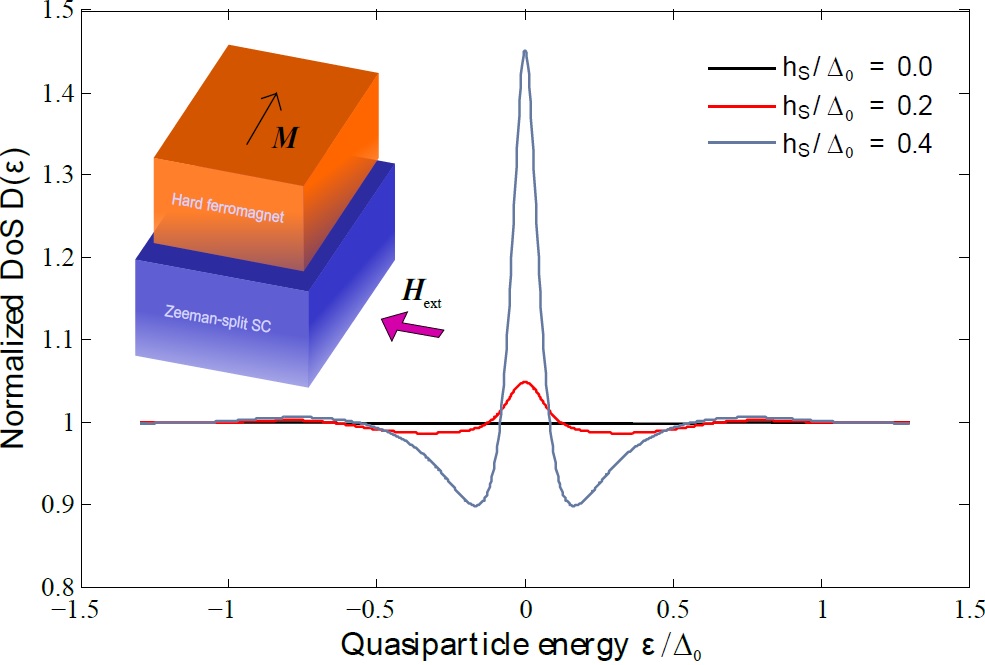}}
\caption{(Color online) Proposed experimental setup: a Zeeman-split superconductor proximity coupled to a magnetically hard ferromagnet. The density of states is obtained at the F/vacuum interface and several values of the exchange field $h_S$ induced by the external field is shown. We have used $T/T_c=0.1$, $h_F/\Delta_0=10$, $\zeta=3$, $d_S/\xi_S=10/15$, $d_F/\xi_S=10/15$. To model inelastic scattering, we added a small imaginary $\i\delta$ to the quasiparticle energies with $\delta/\Delta_0 = 10^{-4}.$}
\label{fig:model} 
\end{figure}

In order to prove the existence of odd-frequency pairing in Zeeman-split superconductors, in spite of the fully suppressed DoS, we propose a simple experiment where STM/conductance measurements are performed on a bilayer consisting of a thin-film superconductor and hard ferromagnet. More specifically, the anisotropy field in the ferromagnet should be of order $\sim 2$ Tesla so that application of a weak magnetic field of a few hundred mT would not alter its orientation. The system is shown in the inset of Fig. \ref{fig:model}. Without any applied field, the proximity effect is efficiently suppressed in the ferromagnet due to its exchange field $h_F$ and no spectroscopic alteration should be present at the end of the F layer. When a weak magnetic field is applied, the ferromagnet remains invariant whereas the superconductor, due to its small thickness, becomes Zeeman-split. If odd-frequency pairing is now present in the superconductor, this must give rise to a long-ranged proximity effect mediated by spin-1 triplet correlations due to the fact that the applied field and the magnetization in the ferromagnet are not aligned \cite{bergeret}. The evidence of this would be a substantially enhanced DoS at the Fermi level in the ferromagnet, even at its end since the spin-1 triplets may propagate over distances $\sim$ 100 nm or more. 

To determine if this is the case, we have solved the full Usadel equation \cite{usadel} for the S/F bilayer numerically without any assumptions of a weak proximity effect. Thus, we are now considering a proximity effect system between a Zeeman-split superconductor and a ferromagnet.  Importantly, we have done so in a self-consistent manner as is required due to the small thickness of both the superconducting and ferromagnetic region, thus taking into account the spatial profile of the superconducting gap and the pair-breaking effect of the interface (parameters used for the numerical simulation are shown in the figure caption). The equation system that needs to be solved in order to achieve this is conveniently expressed using the so-called Ricatti-parametrization \cite{ricatti} which was worked out in the diffusive regime described by the Usadel equation in Ref. \cite{ricusa}, where the Green's function is written in terms of two 2$\times$2 spin matrices $\gamma$ and $\tilde{\gamma}$: 
\begin{align}
\hat{g} = \begin{pmatrix} 
N(1+\gamma\tilde{\gamma}) & 2N\gamma \\
-2\tilde{N}\tilde{\gamma} &-\tilde{N}(1+\tilde{\gamma}\gamma)\\
\end{pmatrix}
\end{align}
with $N=(1-\gamma\tilde{\gamma})^{-1}$ and $\tilde{N} = (1-\tilde{\gamma}\gamma)^{-1}$. This representation automatically satisfies the requirement of $\check{g}^2=\hat{1}$, as can be checked by direct multiplication. The Usadel equation for $\gamma$ reads
\begin{align}\label{eq:ricatti}
D[\partial_x^2\gamma &+ 2(\partial_x\gamma)(1-\tilde{\gamma}\gamma)^{-1}\tilde{\gamma}(\partial_x\gamma) + 2\i\varepsilon\gamma + \i\boldsymbol{h}\cdot(\boldsymbol{\sigma}\gamma - \gamma\boldsymbol{\sigma}^*) + \Delta(\sigma_y-\gamma\sigma_y\gamma) = 0,
\end{align}
whereas the corresponding equation for $\tilde{\gamma}$ is obtained from Eq. (\ref{eq:ricatti}) by complex conjugating and letting $\varepsilon \to -\varepsilon$. To obtain $\{\gamma,\tilde{\gamma}\}$ in the Zeeman-split superconductor and the ferromagnetic region, we also need boundary conditions which at the outer edges of the system simply read $\partial_x\gamma=\partial_x\tilde{\gamma}=0$ whereas at the interface between them we use the Kupriyanov-Lukichev boundary condition \cite{KL} $2\zeta_jL_j \hat{g}_j\partial_x\hat{g}_j = [\hat{g}_S,\hat{g}_F]$, where $j=\{S,F\}$. Here, $\hat{g}_{S,F}$ is the Green's function in the S and F region, respectively, $L_j$ is the width of region $j$ while $\zeta_j=R_B/R_j$ with $R_B$ being the barrier resistance and $R_j$ the resistance of region $j$. Finally, the above equations need to be solved together with the self-consistency equation for the superconducting order parameter: $\Delta = NV \text{Tr}\{\hat{\rho}\int \text{d}\varepsilon \hat{g}^K\}$. Here, $\hat{g}^K$ is the Keldysh part of the Green's function, $\hat{\rho}$ is a 4$\times$4 matrix that projects out the singlet anomalous Green's function component in $\hat{g}^K$, while $N$ and $V$ denote the DoS at the Fermi level in the normal-state of the superconductor and the net attractive interaction between the electrons  This can be done in an iterative fashion, where one starts with an initial guess for the superconducting order parameter, solves the Usadel equations, compute a new profile for $\Delta$, and so on until self-consistency is achieved (meaning that the solutions do not change more than some specified tolerance value). In our simulations, the tolerance was set to $10^{-4}$ (corresponding to $\sim0.1\%$ of the bulk gap) with a weak coupling constant $NV = 0.2$ and self-consistency was typically achieved after 7-8 iterations.

The results are shown in Fig. \ref{fig:model} for the cases of no applied field and fields inducing an exchange splitting of $h_S=0.2\Delta_0$ and $h_S=0.4\Delta_0$, respectively. As predicted, a zero-energy peak emerges at non-zero field whereas the DoS is completely flat without it. An advantage of our setup compared to other proximity structures that have been studied is that the zero-energy peak is large even when the ferromagnet has a strong exchange field $h_F\gg\Delta_0$ [we have chosen $h_F/\Delta_0=10$ for Fig. \ref{fig:model}], a property which usually renders the DoS almost completely featureless. This makes the effect we predict clearly discernable, whereas previous experiments that have looked for zero-energy alterations of the DoS in S/F bilayers have obtained changes of order 1\% of the normal-state DoS \cite{kontos_prl_01, experiment_oddfreq}. Another advantage is that the effect can be turned on and off with a weak magnetic field, since our setup does not require any change of the magnetization direction in the ferromagnet which can be more difficult to achieve if a considerable magnetic anisotropy must be overcome.

\section*{Discussion}

The mixing of conventional singlet and odd-frequency triplet superconductivity in the presence of a magnetic field has been noted before \cite{matsumoto_jpsj_12}, whereas our manuscript is concerned with how odd-frequency pairing is manifested in experimentally accessible quantities. We have shown that a fully gapped DoS in a superconductor / ferromagnet system does not rule out odd-frequency triplet correlations, in contrast to previous predictions on how this type of superconductivity leads to an enhancement of the low-energy density of states \cite{braude, theory_oddfreq}. We mention here also the controversy regarding the thermodynamic stability of an \textit{intrinsic bulk odd-frequency} superconducting state which has been discussed in Refs. \cite{solenov_prb_09, kusunose_jpsj_11, fominov_prb_15}. In contrast, we have considered \textit{induced} triplet pairing in a conventional $s$-wave superconductor which is a stable phenomenon that is well-known to exist from earlier studies of proximity structures. Our findings could have important consequences regarding experimental identification of unconventional superconducting order in a large range of different physical systems such as noncentrosymmetric superconductors and proximity-induced superconductivity in topological insulators due to the appearance of odd-frequency pairing in such a setting. In particular, we have shown that one of the usual hallmark signatures of unconventional superconducting order, the zero-bias conductance peak, can be completely absent even for strong odd-frequency pairing. It could be interesting to explore how antisymmetric spin-orbit coupling influences this, as it has very recently been shown \cite{sol1,sol2} that such interactions strongly influence the spectroscopic signature of triplet correlations.

\section*{Acknowledgements}

 J.L was supported
by the Research Council of Norway, Grants No. 205591 and 216700 and the "Outstanding Academic Fellows" programme at NTNU. J.W.A.R. acknowledges financial support from the Royal Society  ("Superconducting Spintronics") and through a Leverhulme Trust International Network Grant (grant IN-2013-033). The authors thank F. S. Bergeret for discussions.

\section*{Author contributions statement}
J.L. and J.W.A.R. jointly came up with the idea for the project. J.L. performed the analytical and numerical calculations. Both authors contributed to the discussion of the results and the writing of the manuscript.

\section*{Additional information}

\textbf{Competing financial interests} The authors declare no competing financial interests.

\end{document}